\documentclass{article}
\usepackage[T1]{fontenc} 
\usepackage[utf8]{inputenc} 
\usepackage{ismir,amsmath,cite,url}
\usepackage{graphicx}
\usepackage{color}
\usepackage{float}
\usepackage{tabularx}
\usepackage{lineno}
\usepackage{multirow}
\usepackage{booktabs}
\usepackage{algorithm}
\usepackage{algpseudocode}
\usepackage{todonotes}

\title{Skip That Beat: Augmenting Meter Tracking Models for Underrepresented Time Signatures}

\threeauthors
  {Giovana Morais} {MARL\\New York University \\ {\tt giovana.morais@nyu.edu}}
  {Brian McFee} {MARL-CDS\\New York University \\ {\tt brian.mcfee@nyu.edu}}
  {Magdalena Fuentes} {MARL-IDM\\New York University \\ {\tt mfuentes@nyu.edu}}

\def\authorname{G. Morais, B. McFee, and M. Fuentes}

\usepackage[bookmarks=false,pdfauthor={\authorname},pdfsubject={\papersubject},hidelinks]{hyperref}

\newcommand{\multicomment}[1]{}

\begin{document}

\maketitle
\begin{abstract}
Beat and downbeat tracking models are predominantly developed using datasets
with music in 4/4 meter, which decreases their generalization to repertories in
other time signatures, such as Brazilian samba which is in 2/4. In this work, we
propose a simple augmentation technique to increase the representation of time
signatures beyond 4/4, namely 2/4 and 3/4. Our augmentation procedure works by
removing beat intervals from 4/4 annotated tracks. We show that the augmented
data helps to improve downbeat tracking for underrepresented meters while
preserving the overall performance of beat tracking in two different models. We
also show that this technique helps improve downbeat tracking in an unseen samba
dataset.
\end{abstract}
\section{Introduction}\label{sec:introduction}


Current datasets for beat and downbeat tracking are biased towards music in 4/4 meter, which do not adequately address music with different rhythmic structures, such as jazz, which typically 
contains 3/4, 12/8, or even 5/4 meters; classical music, typically featuring 2/4, 3/4, 6/8 meters (among others);
Turkish Aksak (9/8); Cretan leaping dances (2/4), or Brazilian samba, which is in 2/4.

Recent advancements in deep learning have shifted beat and downbeat tracking from traditional signal processing 
methods, which use manually crafted features, to data-driven techniques, such as ~\cite{bockDECONSTRUCTANALYSERECONSTRUCT2020,zhao2022beattransformer,cheng2023transformerbeat,foscarin2024beatthis}. This exacerbated the bias towards 
music in 4/4 because annotating and creating new datasets for
diverse musical meters is both time-consuming and costly. As many culturally specific music genres 
feature meters other than 4/4, this results in a predominance of mainstream annotated musical data \cite{tempobeatdownbeat:book}. 
For example, in \cite{bockDECONSTRUCTANALYSERECONSTRUCT2020} from the 2216 training tracks available, 1120 are in 4/4, 882 do not have beat position annotations, and the remaining 214 spans other 4 time signatures.
This underrepresentation is not a problem for beat tracking, but it affects downbeat tracking as briefly discussed in \cite{krebsDOWNBEATTRACKINGUSING2016}.

 Recent work has explored methods to minimize
the number of required annotations while still enabling models to generalize
across different music styles~\cite{maiaAdaptingMeterTracking2022} and
strategies for selecting which tracks to
annotate~\cite{maiaSelectiveAnnotationFew2024}. While this minimizes the number of tracks and time to annotate, it does not remove the need for annotation.

Another way to alleviate the need to annotate new data is through data
augmentation. Data augmentation involves modifying existing data to improve
model training, such as rotating an image in computer vision tasks or adding noise in audio tasks. 
In audio applications, \cite{mcfee2015SOFTWAREFRAMEWORKMUSICAL} proposes a framework to augment
annotated data, which supports operations such as pitch shift and time stretch,
and evaluate it on instrument recognition. They show that even simple
transformations can lead to improvements in the task, but they should be done
carefully (e.g. avoid deforming the acoustic audio signal too much).
In \cite{bockDECONSTRUCTANALYSERECONSTRUCT2020} the augmentation is done by
calculating the same STFT with different hop lengths, which improves the tempo estimation 
on unseen tempo ranges. Finally, \cite{cheng2023transformerbeat} proposes augmenting the dataset by applying Harmonic-Percussive Source Separation (HPSS), and 
training the model with both the original mel-spectrogram, the harmonic component, and the percussive component.
Inspired by the fact that data augmentation helps models generalize to out-of-domain situations \cite{bockDECONSTRUCTANALYSERECONSTRUCT2020}, we explore these ideas in the context of meter tracking. 

In particular, we propose a novel augmentation procedure that uses existing
annotated data to enhance the representation of underrepresented meters, without
the need for new annotations. 
We evaluate our approach using two models, the
Temporal Convolutional Network (TCN) \cite{bockDECONSTRUCTANALYSERECONSTRUCT2020} and BayesBeat \cite{whiteleyBayesianModellingTemporal2006}. We use a test set with unseen meters but seen music genres, and another test set with an unseen music genre (Brazilian samba) and unseen meter. Our results
demonstrate that our augmentation technique improves downbeat tracking performance while
preserving the overall effectiveness of beat-tracking methods.

\section{Method}

\begin{figure*}[ht]
    \centering
    \includegraphics[width=\linewidth]{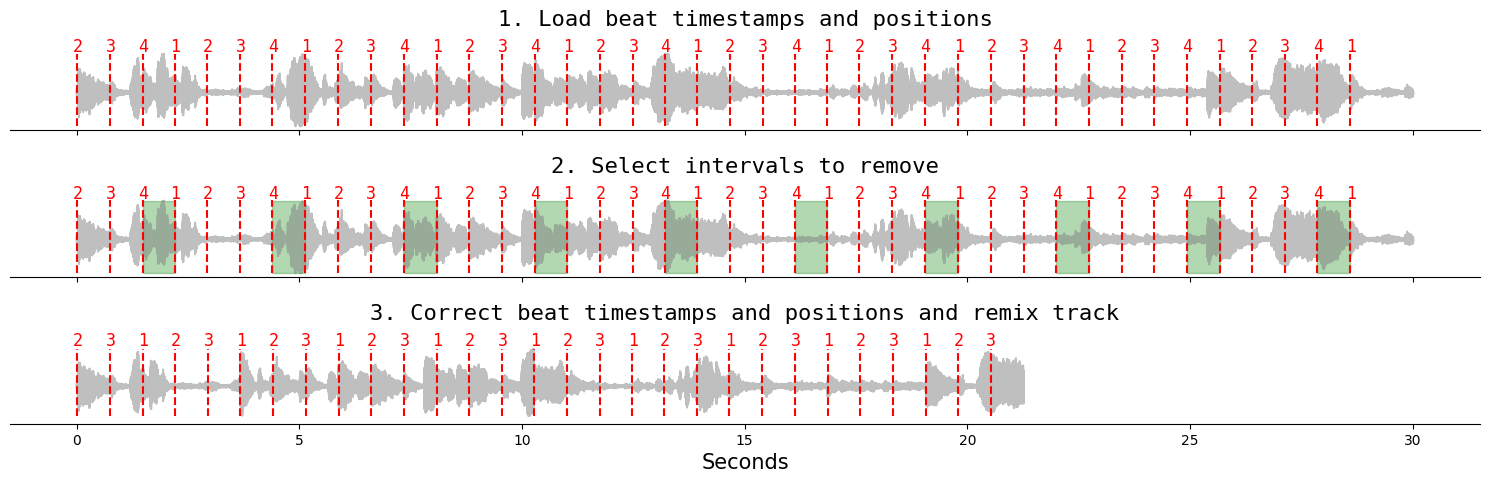}
    \caption{Illustration of the $4/4 \rightarrow 3/4$ augmentation of the GTZAN track \texttt{blues.00009}.
    On the top plot, we have the original
    track with the beat timestamps and positions.
    The middle plot shows the intervals we remove in green. In this case, we remove every
    interval from beat 4 to beat 1. Then, we correct the annotation labels and timestamps and remix the track.
    On the bottom plot, we have the augmented track, with corrected time displacements.}
    \label{fig:augmentation}
\end{figure*}

\subsection{Augmentation Procedure}\label{ssec:augmentation}


Given a track that is in 4/4, let $B = \{b_1, b_2, \cdots\}$ be
the beat annotation timestamps and $P = \{p_1, p_2, \cdots\}$ be the beat
position within the bar, where $p_i \in \{1,2,3,4\}$.
Let $IBI$ be the Inter-Beat Interval of $B$, i.e. $IBI = \Delta B$. We denote
$p^*$ the beat positions we want to keep. For an augmentation $4/4 \rightarrow
2/4,\;p^* = \{1,2\}$, and, similarly, $4/4 \rightarrow 3/4,\; p^* =
\{1,2,3\}$.
Therefore, we define an augmentation as keeping positions $p_i \in p^*$, that is:

\begin{equation}
\hat{B} = \{b_1, b_2, \cdots\}\;\text{if}\;p_i \in p^*
\end{equation}
\begin{equation}
\hat{P} = \{p_1, p_2, \cdots\}\;\text{if}\;p_i \in p^*
\end{equation}
\begin{equation}
	\widehat{IBI} = \{IBI_{i-1} \;\text{if}\;p_i \in p^*\}
\end{equation}

After we remove unwanted beats, we correct the remaining beat positions and any time displacements of the annotations. We do so by taking the first beat timestamp of
$\hat{B}$ and adding the $\widehat{IBI}[i]$ for every new position. We denote the
corrected beat timestamps as $\bar{B}$:
\begin{equation}
	\bar{B} = \{\bar{B}[i-1] + \widehat{IBI}[i]\}
\end{equation}
where $\bar{B}[0] = \hat{B}[0]$. Our augmentation procedure is illustrated in~\figref{fig:augmentation}, and its pseudo-code is provided in Algorithm~\ref{alg:augmentation}. While we do not use the remaining values of $\hat{B}$,
we still calculate it because beat timestamps and beat positions do not necessarily start 
at $p_i = 1$ so it is incorrect to just do $\bar{B}[0] = B[0]$ (see~\figref{fig:augmentation} for an example).

Once we have the corrected new annotations, we use librosa's remix
function\footnote{\url{https://librosa.org/doc/0.10.2/generated/librosa.effects.remix.html}}
to synthesize the augmented signal. This function receives the original audio and
the beat intervals we wish to keep, mapping the interval boundaries to the
closest zero-crossing in the signal to avoid discontinuities and concatenating
them.

We note that, as of now, the beat position removed is fixed and we do
not remove 1st and 2nd beats to keep musical information related to the
downbeat, but these can be further explored in the future. 
It is possible to use the same procedure to augment 4/4 tracks to other time signatures, such as 5/4, but instead of removing beats, one would repeat them. For example, we could repeat the third beat. We leave the exploration of these scenarios for future work.

\begin{algorithm}
\caption{Augmentation Procedure}\label{alg:augmentation}
\begin{algorithmic}
\State $B \gets \{b_1, b_2, \cdots\}$\Comment{beat timestamp annotations}
\State $P \gets \{p_1, p_2, \cdots\}$\Comment{beat positions annotations}
\State $IBI \gets \{0\}$
\State $IBI \gets IBI.\text{append}(\Delta B)$
\State keep $\gets \{\}$ \Comment{indices of positions we want to keep}
\For{$i$ in length($B$)}
    \If{$p \in p^*$}
		\State \text{keep} $\gets\; \text{keep}.\text{append}(i)$
    \EndIf
\EndFor
\State $\widehat{IBI} \gets IBI[keep]$
\State $\hat{P} \gets P[keep]$
\State $\bar{B} \gets \{B[keep[0]]\}$
\For{i in range$(1, \text{length}(\widehat{IBI}))$}
	\State $\bar{B} \gets \bar{B}.\text{append}(\bar{B}[i-1] + \widehat{IBI}[i])$
\EndFor
\end{algorithmic}
\end{algorithm}

\begin{figure}[ht]
    \centering
    \includegraphics[width=\linewidth]{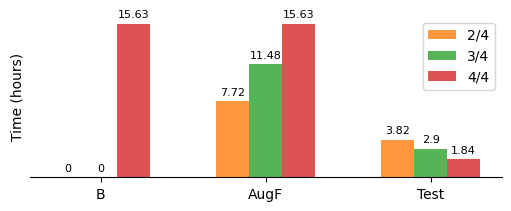}
    \caption{Training and test data distributions by time signature.}
    \label{fig:dataset_distribution}
\end{figure}

\subsection{Datasets}\label{ssec:datasets}

To create our augmented datasets, we use data from the
Beatles~\cite{daviesEvaluationMethodsMusical},
GTZAN~\cite{tzanetakisMusicalGenreClassification2002},
and the RWC Classical and Jazz~\cite{gotoRWCMusicDatabase2002} datasets.
The GTZAN dataset consists of 1000 tracks of 30s each and spans 10 different
genres. The rhythmic annotations, such as tempo, swing ratio, and beat/downbeat
positions, were provided by~\cite{marchandGTZANRhythmExtendingGTZAN2015} with
metrical annotations added by~\cite{quintonExtractionMetricalStructure2015}.
The Beatles dataset consists of 179 songs
from the 12 studio albums from The Beatles. RWC Classical and RWC Jazz consist
of 50 tracks each with variate lengths. All datasets have beat and downbeat annotations, among others. We discard tracks without beat annotations.

For the four aforementioned datasets, we have tracks with no meter annotations.
We infer the meter directly from the beat positions by first calculating the difference between consecutive $p_i$ and identifying the negative differences, 
which indicate the transition between bars, e.g. $[1,2,3,4,1]$ becomes $[1,1,1,-3]$. We count how often these transitions occur at specific beat positions $p_i$. 
We consider the numerator of the meter the most frequent transition point. 
The denominator is always $4$ for the tracks we are using in this work.
This method does not account for tracks with internal time signature changes but gives
us a good approximation of the dominant track meter. 

While GTZAN and
Beatles are more concentrated in 4/4, RWC Classical and Jazz have
better representation on 2/4 and 3/4 time signatures. Still, from the 28.41
hours of audio (1283 tracks) from the datasets, 17.48 hours (1096 tracks) are
classified as 4/4, 2.9 hours (87 tracks) are classified as 3/4, 3.82 hours
(59 tracks) as 2/4 and the other 1.21 hours (41 tracks) are spanned between
the other 5 time signatures. In this work, we focus on 2/4, 3/4, and 4/4 meters, 
leaving the remaining meters for future work. 

These datasets give us a reasonable amount of tracks to train and test the models 
with the different time signatures. 
We create two different datasets to train the models:

\begin{description}
    \item[Baseline (B)]{Randomly selected 993 original 4/4 tracks, which stands for 80\% of the whole dataset.}
	\item[Augmented Full (AugF)]{Same tracks as \textbf{B} plus
		their respective augmentations in 2/4 and 3/4. This is
		the largest dataset between the three, totaling 2979 tracks.}
\end{description}

The test split is fixed for the three experiments, has 20\% of the total tracks and consists only of original
tracks. The split has the 2/4 and 3/4 tracks, plus the remaining 4/4 tracks. As a second and unseen test set, we use the acoustic mixtures 
from the Brazilian Rhythmic Instrument Dataset, BRID~\cite{maiaNovelDatasetBrazilian2018}, 
which consists of 93 short tracks of samba (samba-enredo and partido alto).
Tempo varies from track to track, but not within the track. All the tracks are in 2/4.

Figure~\ref{fig:dataset_distribution} shows the time signature distribution of the data. 
We note here the TCN methods use 20\% of the train split as a validation set, while 
BayesBeat uses the whole set as training.


\subsection{Meter Tracking Models}\label{ssec:models}
To evaluate our augmentation strategy, we train two different meter-tracking
models: the TCN and BayesBeat. In
previous works, the two models were shown to be capable of generalizing to
different music corpora~\cite{maiaAdaptingMeterTracking2022,holzapfelTRACKINGODDMETER2014}.

The TCN was first proposed to beat
tracking in~\cite{daviesTemporalConvolutionalNetworks2019}, and then
expanded to simultaneously estimate beat, downbeat, and tempo
in~\cite{bockDECONSTRUCTANALYSERECONSTRUCT2020}.
We use the open-source implementation provided in~\cite{tempobeatdownbeat:book}.
The TCN outputs beat and downbeat activations (i.e. likelihood of beats and downbeats respectively), which are typically further processed using a Dynamic Bayesian Network (DBN) for temporal consistency. 

According to~\cite{bockJointBeatDownbeat2016}, the DBN is good at dealing with
ambiguous observations and finds the global best state sequence given these
observations. \cite{foscarin2024beatthis} argues that the DBN performs well on most pieces commonly used to train and evaluate beat tracking systems because those tracks
have common features (e.g., stable tempo, meter in either 3/4 or 4/4) and that the DBN is most likely to mispredict more challenging data. 
In preliminary experiments, we observed that the DBN would improve overall results by 5\% (i.e. similar to the BayesBeat), but as argued in \cite{foscarin2024beatthis} it would obscure the effects of the augmentation. To better analyze the augmentation effect in the learning of the model 
we replace the DBN post-processing 
with the adaptive thresholding peak-picking method proposed
by~\cite{nieto2018systematicexploration}\footnote{We use the implementation
provided in
\url{https://www.audiolabs-erlangen.de/resources/MIR/FMP/C6/C6S1_PeakPicking.html}}.
This method has a moving window that compares peaks with the median threshold.
We refer to it as \textit{TCN-PP}. 
Finally, the implementation of the TCN in \cite{tempobeatdownbeat:book} contains tempo augmentation. We removed this step to make the comparison with
the BayesBeat fairer.

The second model we use is
BayesBeat~\cite{whiteleyBayesianModellingTemporal2006}, a statistical method
that tracks beat, downbeat, tempo, meter, and rhythmic patterns. We use the MATLAB
implementation available on GitHub\footnote{\url{https://github.com/flokadillo/bayesbeat}}.
In comparison with the TCN, BayesBeat has fewer parameters and trains faster. For this model,
we use the default parameters, i.e. two frequency bands (low and high) and one
rhythmic pattern. We refer to this model as \textit{BB}.

\section{Experiments and Results}

\begin{table*}[]
\centering
\begin{tabular}{ccccccc|ccccc}
 &  & \multicolumn{5}{c|}{Beat} & \multicolumn{5}{c}{Downbeat} \\ \cline{3-12}
 &  & F1 & CMLt & CMLc & AMLt & AMLc & F1 & CMLt & CMLc & AMLt & AMLc \\ \hline
\multicolumn{1}{c|}{\multirow{3}{*}{BB}} & \multicolumn{1}{c|}{B} & 0.70 & 0.53 & 0.46 & 0.76 & 0.61 & 0.41 & 0.31 & 0.29 & 0.49 & 0.45 \\
\multicolumn{1}{c|}{} & \multicolumn{1}{c|}{AugF} & 0.71 & 0.54 & 0.47 & 0.78 & 0.63 & 0.49 & 0.38 & 0.35 & 0.60 & 0.54 \\ \hline
\multicolumn{1}{c|}{\multirow{3}{*}{TCN-PP}} & \multicolumn{1}{c|}{B} & 0.72 & 0.42 & 0.27 & 0.63 & 0.36 & 0.36 & 0.00 & 0.00 & 0.29 & 0.11 \\
\multicolumn{1}{c|}{} & \multicolumn{1}{c|}{AugF} & 0.74 & 0.47 & 0.33 & 0.64 & 0.39 & 0.36 & 0.01 & 0.00 & 0.23 & 0.11 \\
\end{tabular}%
\caption{Average results for the models across all time signatures.}
\label{tab:general_results}
\end{table*}

\subsection{Setup}

We start by loading all 4/4 tracks from the datasets discussed in
Section~\ref{ssec:datasets}. We use \texttt{mirdata}~\cite{fuentes_2023_10070589} to
load the tracks' annotations. We discard tracks without annotations. 
Then, we apply the augmentation
procedure described in Section~\ref{ssec:augmentation} to all 4/4 tracks.

For every model, we evaluate the F-measure and continuity metrics CMLt, AMLt, CMLc, and AMLc (Correct and 
Allowed Metrical Level with and without continuity required, respectively).  We use the implementation 
provided in \texttt{madmom}\footnote{\url{https://github.com/CPJKU/madmom}}~\cite{madmom}.

In our GitHub repository\footnote{\url{https://github.com/giovana-morais/skip_that_beat}}, we have made
available the augmentation and experiments code.

\subsection{Results}

\tabref{tab:general_results} shows the average performance of the models trained on three 
different datasets. Overall, the beat metrics remained stable across models and datasets, with 
a 2\% improvement for the TCN-PP model on the \textbf{AugF} dataset, mainly due to improvements in 
the 3/4 and 4/4 time signatures.
For downbeat, we see that the BayesBeat model benefited from the \textbf{AugF} dataset with an 
8\% overall improvement. The TCN's performance does not seem to change on average, but when looking at a breakdown of the results per meter we observe some differences, as discussed below.


\subsubsection{Results by Time Signature}

\begin{figure}[ht]
    \centering
    \includegraphics[width=\linewidth]{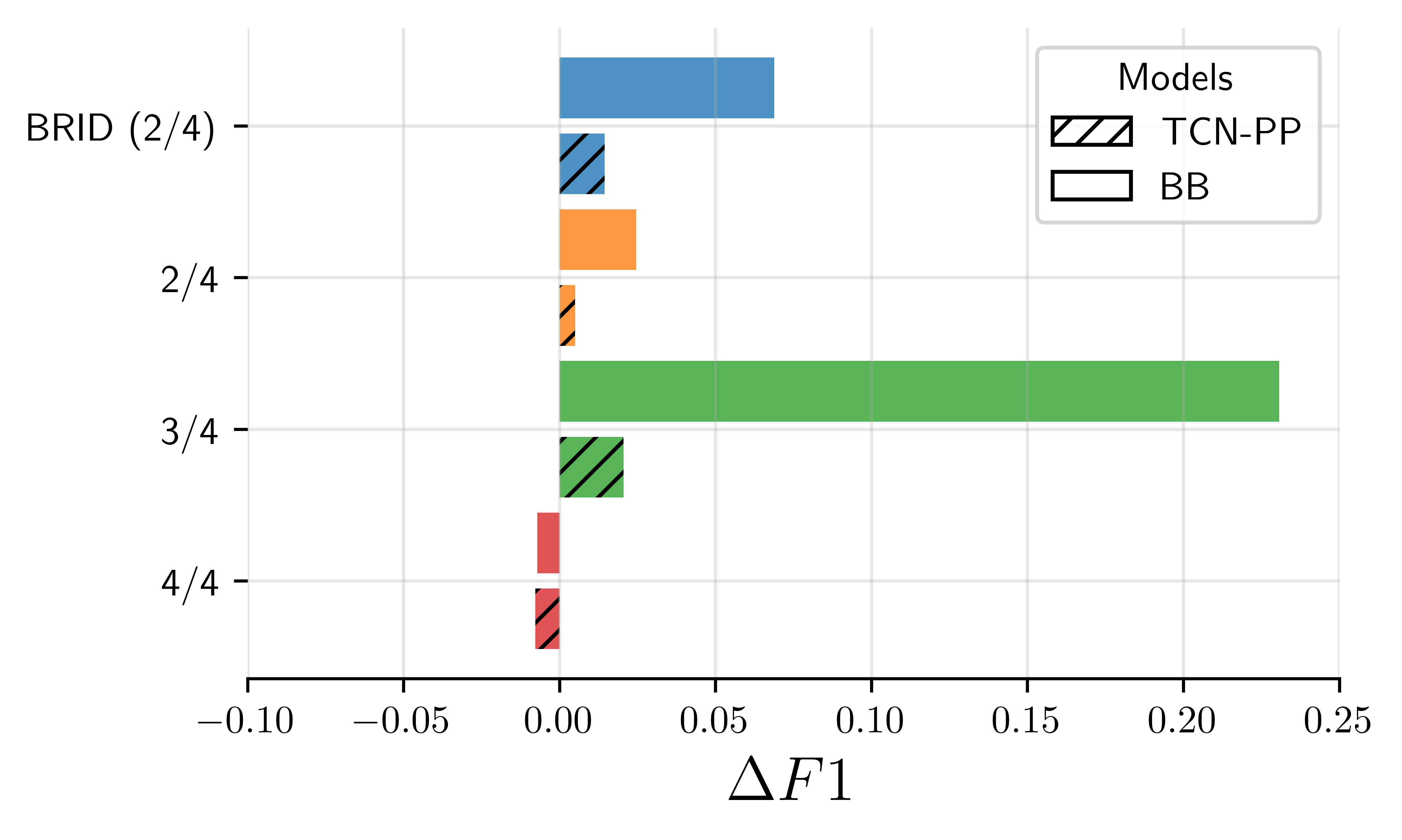}
	\caption{Relative F-measure for downbeat tracking on test set and BRID.
		The bars represent the deviation from the baseline F-measure. Hatched
		bars represent TCN-PP results and non-hatched bars represent BB.}
    \label{fig:downbeat_results}
\end{figure}

Figure \ref{fig:downbeat_results} shows the models' performance broken down by meter in the test set, for the two augmented training datasets. We see an improvement across models in 2/4 and 3/4, with the biggest improvement seen in downbeat tracking for 3/4 tracks for the BB model. We hypothesize that this is due 
to the difference between the time signatures. While 2/4 and 4/4 can be considered similar, 3/4 has a more 
distinct pattern. By inspecting examples, we see that models trained with the 
augmented datasets misclassified 4/4 tracks as 2/4 and the other way around.
Those errors make sense given the perceptual similarity between those meters. Besides the F-measure, \textit{BB} also improves the continuity metrics, as seen in \tabref{tab:general_results}, showing that once the method understands the correct meter, it propagates it through the track.


\subsubsection{Results for BRID}
In~\cite{maiaAdaptingMeterTracking2022} the authors reported results for
a TCN trained only on Western music on the BRID dataset. The 
model had an F-measure of 0.096 and a CMLt of 5.9, demonstrating the difficulty
of generalizing to unseen data. The result had an increase of 70\%
when the authors fine-tuned the model with only 0.67 minutes of annotated BRID data. They argued 
that the difficulty is not only in the meter itself but also in the acoustic properties of the data. 
The authors also report
that BayesBeat can achieve similar downbeat F-measures by being trained from scratch with around 3 minutes of annotated data. We see in~\figref{fig:downbeat_results} that our models show an improvement of 
about 5\% and 15\% with respect to the baseline without annotating new data when evaluated on BRID. The improvement for the TCN method was not as big as the BayesBeat, but in absolute values, the TCN had a better performance than BayesBeat. The augmentation was not sufficient to account for the acoustic properties of the track, which are different from training, but it improves downbeat tracking without any extra annotated data. This augmentation could be used to annotate a few minutes of samba to then retrain the models, as a preprocessing step for the approach in \cite{maiaAdaptingMeterTracking2022}.

As one of its main characteristics, samba has a strong accent on the second beat and the development of contrametric structures~\cite{maiaAdaptingMeterTracking2022}. In our qualitative analysis, we saw that offbeat mistakes were the most common in all models and training data variations, i.e. the strongest beat was classified as the downbeat, even though it was not.

We provide full results and listening examples in the accompanying website\footnote{\url{https://giovana-morais.github.io/skip_that_beat_demo/}}.

\section{Conclusions and Future Work}


In this work, we proposed an augmentation procedure to increase representation in 2/4 and 3/4 time signatures for beat and downbeat tracking. We show that the augmentation approach helps models generalize to unseen meters and datasets but it has limitations for unseen accentuation and rhythmic patterns. The simplicity of the method leaves room for exploration in different directions. For future work, we will explore adding internal meter changes and augmenting different time signatures.
We will also experiment with tracks without voice stems, so the augmentation 
cuts are less perceivable, and explore variations of the removed beat positions, instead of
removing, e.g., the 3rd and 4th beats within a bar.
Finally, it is worth testing the effect of the augmentations on deep learning models
that do not rely on the DBN, such as~\cite{foscarin2024beatthis}, and evaluate whether the improvement and
the generalization to unseen datasets, such as BRID, remains. 

\section{Acknowledgments}

We thank Martín Rocamora for sharing his BayesBeat training scripts with us. We also thank the reviewers for their comments and suggestions. This work was partially supported through the NYU IT High Performance Computing resources, services, and staff expertise.

\bibliography{ISMIRtemplate}

\begin{thebibliography}{10}
\providecommand{\url}[1]{#1}
\csname url@samestyle\endcsname
\providecommand{\newblock}{\relax}
\providecommand{\bibinfo}[2]{#2}
\providecommand{\BIBentrySTDinterwordspacing}{\spaceskip=0pt\relax}
\providecommand{\BIBentryALTinterwordstretchfactor}{4}
\providecommand{\BIBentryALTinterwordspacing}{\spaceskip=\fontdimen2\font plus
\BIBentryALTinterwordstretchfactor\fontdimen3\font minus \fontdimen4\font\relax}
\providecommand{\BIBforeignlanguage}[2]{{%
\expandafter\ifx\csname l@#1\endcsname\relax
\typeout{** WARNING: IEEEtran.bst: No hyphenation pattern has been}%
\typeout{** loaded for the language `#1'. Using the pattern for}%
\typeout{** the default language instead.}%
\else
\language=\csname l@#1\endcsname
\fi
#2}}
\providecommand{\BIBdecl}{\relax}
\BIBdecl

\bibitem{bockDECONSTRUCTANALYSERECONSTRUCT2020}
S.~B{\"{o}}ck and M.~E.~P. Davies, ``{Deconstruct, Analyse, Reconstruct: How to improve Tempo, Beat, and Downbeat Estimation},'' in \emph{Proceedings of the 21th International Society for Music Information Retrieval Conference, {ISMIR} 2020, Montreal, Canada, October 11-16, 2020}, 2020, pp. 574--582.

\bibitem{zhao2022beattransformer}
\BIBentryALTinterwordspacing
J.~Zhao, G.~Xia, and Y.~Wang, ``{Beat Transformer: Demixed Beat and Downbeat Tracking with Dilated Self-Attention},'' in \emph{Proceedings of the 23rd International Society for Music Information Retrieval Conference, {ISMIR} 2022, Bengaluru, India, December 4-8, 2022}, P.~Rao, H.~A. Murthy, A.~Srinivasamurthy, R.~M. Bittner, R.~C. Repetto, M.~Goto, X.~Serra, and M.~Miron, Eds., 2022, pp. 169--177. [Online]. Available: \url{https://archives.ismir.net/ismir2022/paper/000019.pdf}
\BIBentrySTDinterwordspacing

\bibitem{cheng2023transformerbeat}
T.~Cheng and M.~Goto, ``{Transformer-Based Beat Tracking With Low-Resolution Encoder and High-Resolution Decoder},'' in \emph{Proceedings of the 24th International Society for Music Information Retrieval Conference, {ISMIR} 2023, Milan, Italy, November 5-9, 2023}, 2023, pp. 466--473.

\bibitem{foscarin2024beatthis}
F.~Foscarin, J.~Schl{\"{u}}ter, and G.~Widmer, ``{Beat this! Accurate beat tracking without {DBN} postprocessing},'' \emph{CoRR}, vol. abs/2407.21658, 2024.

\bibitem{tempobeatdownbeat:book}
\BIBentryALTinterwordspacing
M.~E.~P. Davies, S.~{Böck}, and M.~Fuentes, \emph{{Tempo, Beat and Downbeat Estimation}}, Nov. 2021. [Online]. Available: \url{https://tempobeatdownbeat.github.io/tutorial/intro.html}
\BIBentrySTDinterwordspacing

\bibitem{krebsDOWNBEATTRACKINGUSING2016}
F.~Krebs, S.~B{\"{o}}ck, M.~Dorfer, and G.~Widmer, ``{Downbeat Tracking Using Beat Synchronous Features with Recurrent Neural Networks},'' in \emph{Proceedings of the 17th International Society for Music Information Retrieval Conference, {ISMIR} 2016, New York City, United States, August 7-11, 2016}, 2016, pp. 129--135.

\bibitem{maiaAdaptingMeterTracking2022}
L.~S. Maia, M.~Rocamora, L.~W.~P. Biscainho, and M.~Fuentes, ``Adapting meter tracking models to latin american music,'' in \emph{Proceedings of the 23rd International Society for Music Information Retrieval Conference, {ISMIR} 2022, Bengaluru, India, December 4-8, 2022}, 2022, pp. 361--368.

\bibitem{maiaSelectiveAnnotationFew2024}
------, ``{Selective Annotation of Few Data for Beat Tracking of Latin American Music Using Rhythmic Features},'' \emph{Trans. Int. Soc. Music. Inf. Retr.}, vol.~7, no.~1, pp. 99--112, 2024.

\bibitem{mcfee2015SOFTWAREFRAMEWORKMUSICAL}
B.~McFee, E.~J. Humphrey, and J.~P. Bello, ``{A Software Framework for Musical Data Augmentation},'' pp. 248--254, 2015.

\bibitem{whiteleyBayesianModellingTemporal2006}
N.~Whiteley, A.~T. Cemgil, and S.~J. Godsill, ``{Bayesian Modelling of Temporal Structure in Musical Audio},'' in \emph{{ISMIR} 2006, 7th International Conference on Music Information Retrieval, Victoria, Canada, 8-12 October 2006, Proceedings}, 2006, pp. 29--34.

\bibitem{daviesEvaluationMethodsMusical}
M.~E.~P. Davies, N.~Degara, and M.~D. Plumbley, ``Evaluation {{Methods}} for {{Musical Audio Beat Tracking Algorithms}}.''

\bibitem{tzanetakisMusicalGenreClassification2002}
G.~Tzanetakis and P.~R. Cook, ``Musical genre classification of audio signals,'' \emph{{IEEE} Trans. Speech Audio Process.}, vol.~10, no.~5, pp. 293--302, 2002.

\bibitem{gotoRWCMusicDatabase2002}
M.~Goto, H.~Hashiguchi, T.~Nishimura, and R.~Oka, ``{RWC} music database: Popular, classical and jazz music databases,'' in \emph{{ISMIR} 2002, 3rd International Conference on Music Information Retrieval, Paris, France, October 13-17, 2002, Proceedings}, 2002.

\bibitem{marchandGTZANRhythmExtendingGTZAN2015}
U.~Marchand, Q.~Fresnel, and G.~Peeters, ``{{GTZAN-Rhythm}}: {{Extending}} the {{GTZAN Test-Set}} with {{Beat}}, {{Downbeat}} and {{Swing Annotations}},'' Oct. 2015.

\bibitem{quintonExtractionMetricalStructure2015}
E.~Quinton, C.~Harte, M.~Sandler, and C.~Shannon, ``{Extraction of Metrical Structure from Music Recordings},'' in \emph{Proceedings of the 18th {{International Conference}} on {{Digital Audio Effects}} ({{DAFx-15}})}, Trondheim, Norway, 2015.

\bibitem{maiaNovelDatasetBrazilian2018}
P.~Tomaz, L.~Maia, M.~Fuentes, M.~Rocamora, L.~Biscainho, M.~{da Costa}, and S.~Cohen, ``{A Novel Dataset of Brazilian Rhythmic Instruments and Some Experiments in Computational Rhythm Analysis},'' in \emph{{Congreso Latinoamericano de la AES 2018, Montevideo, Uruguay, September 24-26, 2018}}, Sep. 2018.

\bibitem{holzapfelTRACKINGODDMETER2014}
A.~Holzapfel, F.~Krebs, and A.~Srinivasamurthy, ``{Tracking the "Odd": Meter Inference in a Culturally Diverse Music Corpus},'' in \emph{Proceedings of the 15th International Society for Music Information Retrieval Conference, {ISMIR} 2014, Taipei, Taiwan, October 27-31, 2014}, H.~Wang, Y.~Yang, and J.~H. Lee, Eds., 2014, pp. 425--430.

\bibitem{daviesTemporalConvolutionalNetworks2019}
E.~P. MatthewDavies and S.~B{\"{o}}ck, ``Temporal convolutional networks for musical audio beat tracking,'' in \emph{27th European Signal Processing Conference, {EUSIPCO} 2019, {A} Coru{\~{n}}a, Spain, September 2-6, 2019}.\hskip 1em plus 0.5em minus 0.4em\relax {IEEE}, 2019, pp. 1--5.

\bibitem{bockJointBeatDownbeat2016}
S.~B{\"{o}}ck, F.~Krebs, and G.~Widmer, ``{Joint Beat and Downbeat Tracking with Recurrent Neural Networks},'' pp. 255--261, 2016.

\bibitem{nieto2018systematicexploration}
O.~Nieto and J.~P. Bello, ``{Systematic Exploration of Computational Music Structure Research},'' in \emph{Proceedings of the 17th International Society for Music Information Retrieval Conference, {ISMIR} 2016, New York City, United States, August 7-11, 2016}, 2016, pp. 547--553.

\bibitem{fuentes_2023_10070589}
M.~Fuentes, R.~Bittner, G.~{Plaja-Roglans}, G.~Cort{\`e}s, T.~Khandelwal, H.~Palan, M.~Miron, D.~Zasukha, C.~Thom{\'e}, G.~Morais, F.~Papaleo, P.~Ramoneda, J.~Arruti, and M.~Rocamora, ``Mirdata 0.3.8,'' Zenodo, Nov. 2023.

\bibitem{madmom}
S.~B{\"o}ck, F.~Korzeniowski, J.~Schl{\"u}ter, F.~Krebs, and G.~Widmer, ``{madmom: a new Python Audio and Music Signal Processing Library},'' in \emph{Proceedings of the 24th ACM International Conference on Multimedia}, Amsterdam, The Netherlands, 10 2016, pp. 1174--1178.

\end{thebibliography}

\end{document}